\documentclass[aps,prb,twocolumn,amsfonts,amsmath,amssymb,superscriptaddress]{revtex4}
\usepackage{graphicx}
\usepackage{dcolumn}
\usepackage{bm}

\usepackage{colordvi}
\usepackage{color}

\begin{document}

\title{Anisotropic spin relaxation revealed by resonant spin amplification in
(110) GaAs quantum wells}

\author{M. \ Griesbeck}
\affiliation{Institut f\"ur Experimentelle und Angewandte Physik,
Universit\"at Regensburg, D-93040 Regensburg, Germany}
\author{M.M.\ Glazov}
\email{glazov@coherent.ioffe.ru}
\affiliation{Ioffe Physical-Technical Institute, Russian Academy of Sciences, 194021 St. Petersburg, Russia}
\author{E.Ya.\ Sherman}
\affiliation{Department of Physical Chemistry, The University of the Basque Country UPV/EHU, 48080 Bilbao, Spain
\\ IKERBASQUE Basque Foundation for Science, Alameda Urquijo 36-5, 48011, Bilbao, Bizkaia, Spain}
\author{D.\ Schuh}
\affiliation{Institut f\"ur Experimentelle und Angewandte Physik,
Universit\"at Regensburg, D-93040 Regensburg, Germany}
\author{W.\ Wegscheider}
\affiliation{Solid State Physics Laboratory, ETH Zurich, 8093 Zurich, Switzerland}
\author{C.\ Sch\"uller}
\affiliation{Institut f\"ur Experimentelle und Angewandte Physik,
Universit\"at Regensburg, D-93040 Regensburg, Germany}
\author{T.\ Korn}
\email{tobias.korn@physik.uni-regensburg.de}
\affiliation{Institut
f\"ur Experimentelle und Angewandte Physik, Universit\"at
Regensburg, D-93040 Regensburg, Germany}
\date{\today}

\begin{abstract}
We have studied  spin dephasing in a high-mobility two-dimensional
electron system (2DES), confined in a GaAs/AlGaAs quantum well grown in the  [110] direction, using the
resonant spin amplification (RSA) technique. From the characteristic
shape of the RSA spectra, we are able to extract the spin dephasing
times (SDT) for electron spins aligned along the growth direction or
within the sample plane, as well as the $g$ factor. We observe a
strong anisotropy in the spin dephasing times.
While the in-plane SDT remains almost constant as the
temperature is varied between 4~K and 50~K, the out-of-plane SDT shows
a dramatic increase at a temperature of about 25~K and reaches values
of about 100~ns.   The SDTs at 4~K can be further increased by
  additional, weak above-barrier illumination. The origin of this unexpected behavior
is discussed, the  SDT enhancement is attributed to the redistribution of charge carriers  between the electron gas and remote donors.
\end{abstract}
\pacs{75.40.Gb 85.75.-d 73.61.Ey}

\maketitle
\section{Introduction}
Two-dimensional electron systems based on the GaAs/AlGaAs materials
are promising candidates for semiconductor
spintronics~\cite{Fabian04,WuReview} devices.
They offer very high electron mobilities and allow one
to manipulate the spin orientation by electric fields~\cite{Harley03}
via the Rashba spin-orbit interaction (SOI)~\cite{Rashbafield}. For
structures grown along the [110] crystallographic direction, spin
dephasing via the Dyakonov-Perel (DP) mechanism~\cite{DP} is strongly
suppressed for growth-axis-oriented spins~\cite{DP110,Ohno99_1}, while
it remains active for other spin orientations~\cite{Oestreich04_1}. A similar anisotropy of the spin dephasing arises in [001]-grown structures for equal strength of Rashba and Dresselhaus fields~\cite{Golub99,liu:112111,stich:073309,Korn20081542,Korn_ICPS08}. For such structures, however, the suppression of the DP mechanism occurs along one \emph{in-plane} crystallographic orientation.
While all-electrical devices are envisioned for most future
semiconductor spintronics applications, optical spectroscopy
techniques have proven to be very useful for the study of spin
dynamics in direct-gap semiconductor heterostructures, and a variety
of techniques, including time-resolved Faraday rotation
(TRFR)~\cite{Baumberg_Faraday}, Hanle measurements and spin noise
spectroscopy (SNS)~\cite{Oestreich_noise} have been developed.
 A number of experimental groups have studied spin dephasing in
  various (110) grown systems. For nominally undoped quantum wells (QWs), growth-axis
  SDTs of 2$\div$4~ns at room temperature were reported~\cite{Ohno99_1,
    Harley03}, and at low temperatures, using surface acoustic waves to laterally transport optically oriented electrons, SDTs of 18~ns were reached~\cite{PhysRevLett.98.036603}. The temperature dependence of the  SDTs in a (110)
  grown 2DES was studied in Ref.~\onlinecite{Oestreich04_1} using time-resolved
  photoluminescence, yielding  values of the
  growth-axis SDT between 1.8~ns at liquid-helium temperature and
  6.5~ns at 120~K. In optical studies of spin dynamics, the use of
  interband excitation or probing always generates electron-holes
  pairs, and the optically created holes
provide a spin dephasing channel via the Bir-Aronov-Pikus (BAP)
mechanism~\cite{BAP}, hampering the approach to the SDT of an
unperturbed system.The largest value for the growth-axis SDT in a
(110) grown 2DES reported so far, 24~ns,~\cite{oestreich08,Mueller2010} was
determined by the SNS technique in the limit of weak optical probing
of the equilibrium spin dynamics. The large SDT values in (110)-grown 2DES also allowed the observation of hyperfine interaction between nuclei and electron spins in an all-optical nuclear magnetic resonance experiment~\cite{ohno01}.

  Here, we present time-resolved optical studies of a high-mobility (110)-grown 2DES using the resonant spin amplification (RSA) technique~\cite{awschalom98,awschalom99},  a variation of the TRFR technique, which has been successfully
applied to study electron and hole spin dynamics in systems of
  different dimensionality~\cite{awschalom99,ast08,Korn10,Yakovlev_RSA_PRL09}.
We observe SDTs of about 100~ns at low temperatures, exceeding the
previously reported values for free electrons in a 2DES by almost one order of magnitude. We show
that the SDTs extracted from RSA spectra are not limited by the BAP
mechanism and use an optical gating technique to control the 2DES
carrier density and growth-axis symmetry to reach even higher SDT
values.

\section{Sample structure and experimental methods}
\begin{figure}
  \includegraphics[width= 0.5 \textwidth]{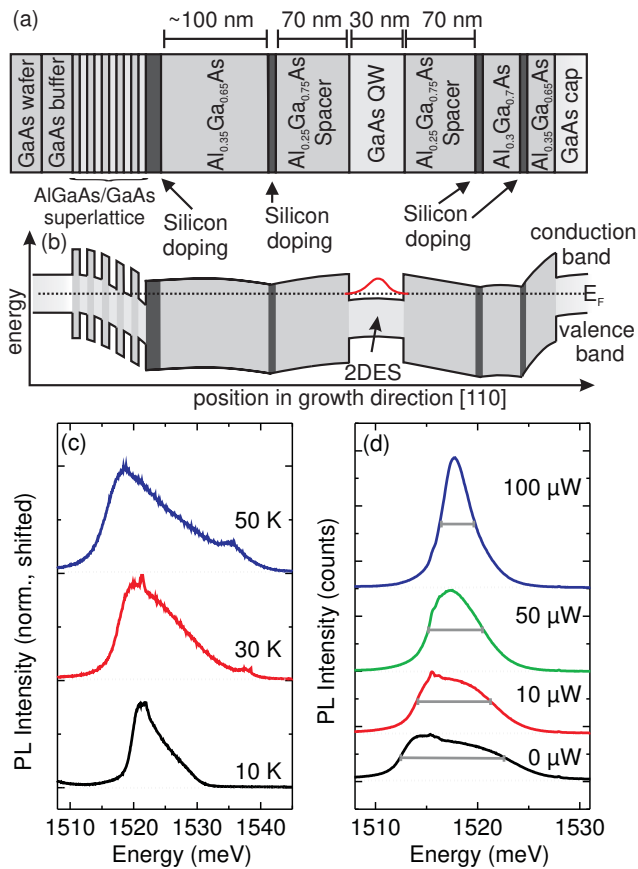}
   \caption{ (a)  Schematic layer structure of the investigated
     sample.  (b) Schematic band structure of the investigated
     sample. (c) PL traces measured at 3 different
     temperatures. (d) PL traces measured at  4~K for different
     excitation densities of above-barrier illumination.}
   \label{Fig1}
\end{figure}

Our sample contains a symmetrically $n$-modulation-doped, 30~nm wide GaAs
QW in which the 2DES resides. It is similar in design to
structures introduced by Umansky \textit{et al.}~\cite{Umansky20091658}. Figure~\ref{Fig1}(a) shows the
  schematic layer structure of the sample: a total of four $n$-doping
  layers are deposited in the barrier material left and right of the
  GaAs QW. While the doping layers far to the left and to the right of the QW mostly
  serve to give flat-band conditions, the two closer doping layers
  provide the charge carriers for the 2DES. These doping layers are embedded between two 2~nm thick layers of AlAs, so that some of the dopant electrons occupy the  $X$ valley states in the AlAs layers and lead to partial screening of the dopant disorder potential~\cite{Roessler2010}. A sketch of the resulting
  band structure is given in Fig.~\ref{Fig1}(b), showing the
  well-defined symmetric confinement of the 2DES in the QW. The nominal carrier density $n=2.7 \times 10^{11}$~cm$^{-2}$  and mobility $\mu=2.3~\times 10^6$~cm$^2$(Vs)$^{-1}$ of our sample were determined at 1.5~K using magnetotransport. In similar structures grown on (001) substrates, even higher carrier mobilities above 18~$\times 10^6$~cm$^2$(Vs)$^{-1}$ were observed at low temperatures, allowing us to study the spin dynamics of electrons on ballistic cyclotron orbits~\cite{Griesbeck09,KornReview}. The
sample  is mounted in vacuum in a
He-flow cryostat during measurements, and the sample temperature is
varied between 4~K and 50~K.  We note that the 2DES electron
temperature [extracted from analysis of the photoluminescence (PL)
lineshape, not shown] is higher than the lattice temperature, and
remains above 15~K even for the lowest sample temperatures,  as the
 high mobility of the sample corresponds to
very inefficient electron-lattice coupling. We utilize a pulsed Ti:sapphire laser system to excite
electrons in the 2DES slightly above the Fermi energy with a
circularly polarized pump pulse, and a time-delayed, linearly
polarized probe pulse from the same laser is used to detect the
growth-axis spin polarization in the 2DES via the spin Kerr
  effect. The laser pulse length is about 2~ps, corresponding to a spectral linewidth of the laser of about 1~meV. The laser energy is kept fixed throughout the temperature-dependent and illumination-dependent measurement series. Pump and probe beams are focused to a spot size of about 50~$\mu$m on the sample using an achromat. For the RSA
measurements, the delay between pump and probe is kept fixed and
adjusted such that the probe pulse arrives about 50~ps \textit{before}
the subsequent pump pulse, and the Kerr signal is recorded as a function of the
applied in-plane magnetic field.
For PL measurements, the pulsed Ti:sapphire laser system is detuned to higher energies to nonresonantly excite electron-hole pairs in the QW. An excitation density of about 0.2~W/cm$^2$ is used for the PL measurement series.  During some of the measurements,
an additional, above-barrier illumination of  the sample
is realized using a green (532~nm) continuous wave ($cw$) laser. The green laser is weakly focused to a spot size of about 1~cm$^2$, which covers the whole sample, to ensure that the above-barrier illumination is homogeneous throughout the sample area probed by the Ti:sapphire laser system.

First, we discuss the shape of the RSA traces observed in our sample,
outline the model and demonstrate how all the relevant spin
  dynamics parameters can be extracted.
  Figure~\ref{Fig2}(a) shows a
typical RSA trace measured on our sample. The signal contains a
  series of peaks corresponding to the commensurability of the spin
  precession period in the external field and the pump pulse repetition
  period $T_{\rm rep}=12$~ns. The peak width and height are related with the spin
  relaxation rates~\cite{Glazov_RSA}.  We clearly see that the RSA
peak centered around zero magnetic field is more pronounced than the
RSA peaks for finite fields, whose heights and widths are equal
in the magnetic-field range investigated.
 This trace shape is a direct evidence of the specific spin-orbit
   field symmetry in almost symmetric (110)-oriented systems
   which is predominantly oriented along the sample growth axis
   $z\parallel [110]$
   (Dresselhaus field). It leads to  fast DP spin relaxation of
   in-plane spin components and slow relaxation of $z$ spin component
   due to either regular or random Rashba
   fields~\cite{GlazovSherman_rev,glazov2009a}. Indeed, for long spin
   relaxation time $T_{zz}$ the $z$ spin component is efficiently
   accumulated at zero magnetic field $B$, resulting in  constructive
   interference of spins created by the train of pump
   pulses, giving rise to a large Kerr signal at negative time delays.
 For applied in-plane magnetic fields, the optically oriented electron
 spins precess into the sample plane, and therefore, more rapid
 dephasing due to the DP mechanism occurs. However, some spin
 polarization remains within the sample during the time between
 subsequent pump pulses, and if the Larmor precession frequency is
 commensurate with the laser repetition rate, constructive
 interference occurs, resulting in weaker and broader maxima.

\section{Theoretical approach}
To obtain a quantitative description of RSA traces we follow
  Ref.~\onlinecite{Glazov_RSA} and derive the following
expression for
 the spin $z$ component as function of $B$ and pump-probe time delay
  ($\Delta t<0$):
\begin{equation}
\label{RSA}
\frac{s_z(\Delta t)}{s_0} = {e^{-(T_{\rm rep}+\Delta t)/{\bar T}}}
\frac{e^{T_{\rm rep}/\bar T}\mathcal C[\tilde \Omega (T_{\rm rep} + \Delta t)]-\mathcal C(\tilde \Omega \Delta t)}
{2[\cosh(T_{\rm rep}/\bar T) - \cos(\tilde \Omega T_{\rm rep})]},
\end{equation}
where $s_0$ is the spin injected by a single pump pulse, ${\bar
  T}^{-1} = (\Gamma_{yy}+\Gamma_{zz})/{2}$, function
$$\mathcal C(\xi) = \cos{\xi} +
[(\Gamma_{yy}-\Gamma_{zz})/2\tilde\Omega] \sin{\xi},$$
 $\Gamma_{ij}$ with $i,j=x,y,z$
($x||[1\bar 1 0],y||[110]$) are    the spin relaxation
rates tensor components and $\tilde \Omega = \sqrt{(g\mu_{\rm B}
  B/\hbar)^2-\Gamma_{yy}^2/4}$ with $g$ being   the electron Land\'{e} factor and
$\mu_{\rm B}$ being   the Bohr magneton is
the electron spin precession frequency. Equation~\eqref{RSA} is derived under assumption that $z$ spin component
	relaxation is driven by regular Rashba field, in which case the spin
	relaxation rates tensor is non-diagonal and spin relaxation time
	$T_{zz} \approx 2/\Gamma_{zz}$, see Ref.~\onlinecite{tarasenko:165317}
	for details. If spin relaxation is determined by random Rashba fields,
	one has $T_{zz} = 1/\Gamma_{zz}$ and $\tilde \Omega = \sqrt{(g\mu_{\rm
	B} B/\hbar)^2-(\Gamma_{yy}-\Gamma_{zz})^2/4}$ in Eq.~\eqref{RSA}.

  It follows from
Eq.~\eqref{RSA} that  $T_{zz}$, the growth-axis SDT, is correlated
with the full width at half maximum (FWHM) of the zero-field RSA peak,
$T_{yy}$, the in-plane SDT, is related to the finite-field RSA-peak
FWHM. The spacing of the RSA peaks is inversely proportional to the
electron $g$ factor. The results of experimental data fitting by
Eq.~\eqref{RSA} are presented in Fig.~\ref{Fig2}(c).

\section{Results and Discussion}
\subsection{Temperature dependence of the spin dephasing}
\begin{figure}
  \includegraphics[width= 0.5 \textwidth]{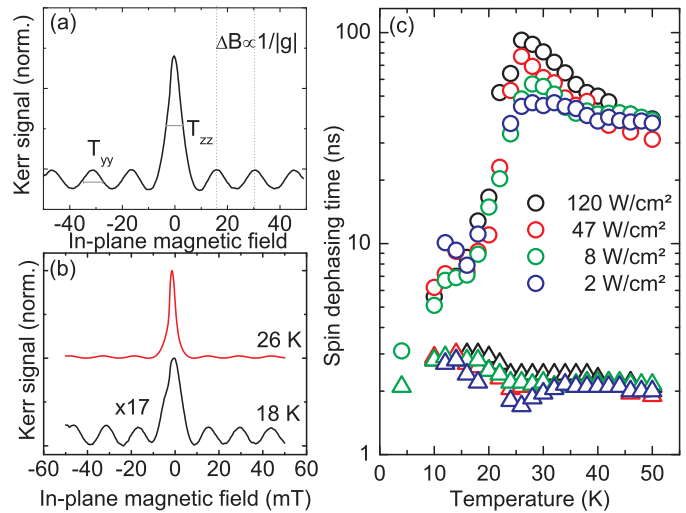}
   \caption{ (a)  Typical RSA trace measured on a high-mobility 2DES
     grown in the [110] direction. The influence of the trace features
     on sample parameters is indicated.  (b) RSA traces measured at
     18~K and 26~K. The data has been normalized and shifted. (c)
     Out-of plane ($T_{zz}$, open circles) and in-plane ($T_{yy}$,
     open triangles) spin dephasing times for different excitation
     densities as a function of temperature.}
   \label{Fig2}
\end{figure}
Let us start with the discussion of the temperature dependence of the SDTs. Figure~\ref{Fig2}(b)  shows
normalized RSA traces measured at 18~K and 26~K.
We note that the zero-field RSA peak drastically increases its
amplitude at the higher temperature, while the finite-field RSA peaks
remain nearly constant. The corresponding values of $T_{zz}$ and
$T_{yy}$ are given in Fig.~\ref{Fig2}(c) in logarithmic
  scale. For the whole range of excitation densities used in our measurements,
 we see a drastic increase of $T_{zz}$ from below 10~ns at low
 temperatures to about 100~ns around 25~K, while $T_{yy}\approx
   2$~ns  shows no markable temperature dependence.
Additionally,  we observe that $T_{zz}$ and $T_{yy}$  become larger as the excitation density is increased.

Now we consider the origin of the spin relaxation times $T_{yy}$ and $T_{zz}$
and their temperature dependence. The relaxation of the in-plane spin components is
well-described by the DP mechanism
resulting from the Dresselhaus SOI
{
\begin{equation}
H_{D}=-\gamma k_{x}\sigma_{z}\left[\langle k_{z}^{2}\rangle+\left(2k_{y}^{2}-k_{x}^{2}\right)\right]/2,
\end{equation}
}
where $\gamma$ is the bulk Dresselhaus coupling constant, $\sigma_{z}$ is the Pauli matrix,
$k_{x},k_{y}$ are in-plane components of the electron wavevector,
and $\langle k_{z}^{2}\rangle\approx \pi^2/w^2$, where $w$ is the QW width.
With the temperature increase, if the electron concentration remains constant, electron-electron
collisions are expected to reduce the DP spin relaxation rate \cite{Glazov04_eeI}.
However, in a complicated system with four remote
dopant layers, a temperature-dependent electron density redistribution as well as ionization
of the donors are expected. This redistribution leads to an increase in the electron concentration
in the 2DES.  To observe this effect, we  perform temperature-dependent
PL measurements. The PL of the 2DES has a characteristic, shark-fin-like shape. It stems from the recombination of electrons from the lowest-lying state in the 2DES up to the Fermi energy, with holes in the valence band. In a 2DES, the Fermi energy is proportional to the carrier density, therefore, the full width at half maximum (FWHM) of the PL may be used to track changes of the local carrier density under excitation conditions  similar to those during the RSA measurements.
As seen in Fig.~\ref{Fig1}(c), the FWHM of the
PL from the 2DES increases with rising
temperature. As the temperature is increased from 4~K to 30~K, the Fermi energy and corresponding electron density of the 2DES almost doubles.
The corresponding increase in the spin precession rate due to the linear and cubic
in the in-plane momentum contributions largely compensates the temperature-induced decrease
in the electron-electron collision time rendering the relaxation time
$T_{yy}$ weakly temperature-dependent.

The Dresselhaus term, however, does not cause relaxation of the
$z$-component.
There are two main origins of the low-temperature value of $T_{zz}$ of the order of
2~ns observed in the experiment [Fig.~\ref{Fig2}(c)].
First, one can expect some ``frozen'' asymmetry $\Delta n$ in the
  electron density
to the left and to the right of
the 2DES due to trapping of carriers, either in the AlAs layers surrounding the doping, or in the spacer layers between
the remote doping and the QW. This asymmetry leads to the Rashba coupling $\alpha_{R}=2\pi\xi e^2\Delta n/\kappa$,
where $\xi=5\times 10^{-2}$ nm$^2$ is the Rashba coefficient for GaAs, $e$ is the electron charge, and $\kappa$ is the dielectric
constant. Using the experimental data demonstrating that at 4K, $T_{zz}\approx T_{yy}$ and
assuming the same DP relaxation mechanism for all spin components, we obtain the condition
$\alpha_{R}\approx \gamma\langle k_{z}^{2}\rangle/2$. Taking into account a broad spread
in experimentally reported values of $\gamma$ from 5 to 28 eV\AA$^{3}$
[Refs. \onlinecite{Pikus1988,Jusserand1995,Miller2003,winkler03,Leyland2007,Koralek2009,Balocchi2011}],
one can estimate that the required asymmetry $\Delta n$ lie in the range between $\sim 0.5\times10^{11}$ and $\sim3\times10^{11}$ cm$^{-2}$.
Second,
the random electric field of the dopants, assuming that they are not fully
screened by the charge carriers in the AlAs layers~\cite{glazov2009a},
leads to a random Rashba field and spin relaxation rate
\begin{equation}
\Gamma_{zz}=\frac{16\pi}{\hbar^{3}}\,\frac{me^{4}\xi^{2}n_{d}k_{F}}{\kappa^{2}R_{d}},
\end{equation}
where $m$ is electron effective mass, $n_d$
is the donor concentration per one side of 2DES, $R_{d}$ is the
distance from 2DES to the dopant layer,
and $k_F$ is the Fermi momentum~\cite{Sherman_RandomRashba}. The
nominal concentration $n_{d}$ of
the order of $5\times10^{12}$~cm$^{-2}$  and the distance to
the 2DES $R_{d}=85$~nm lead to the spin relaxation time of the same
few ns order of magnitude.
With the temperature increase,
the frozen asymmetry disappears, and the charge redistribution of itinerant
electrons and electrons in the vicinity of dopant layers \cite{Roessler2010} switches on the screening of the
random Rashba field leading to the $T_{zz}$ values of the order of 50-100 ns.
\subsection{Excitation density dependence of the spin dephasing}
Next, we focus on the influence of excitation density  on the SDTs. As discussed above,
both, $T_{zz}$ and $T_{yy}$ \textit{increase} with an increase of the
excitation density, in stark contrast to previous measurements on
(110)-grown 2DES~\cite{oestreich08, Voelkl11}, where increasing
excitation (or probing) density lead to reduction of the SDT due to
spin dephasing via the BAP mechanism. To understand this difference,
we need to consider the difference in the experiments: while the spin
noise spectroscopy and Hanle-type experiments utilize $cw$ illumination
of a sample, the RSA technique uses a pulsed laser system. The
remaining spin polarization is probed about 12~ns after pulsed
excitation, when photocarrier relaxation and
recombination taking place on a sub-ns timescale are
complete. Hence, even if the spin relaxation rate is increased while
photocreated holes are present in the sample, the BAP mechanism is
absent for the majority of the measurement time. A trivial explanation for the increase observed
for $T_{zz}$ and $T_{yy}$ with the excitation density is a reduction of the single electron
momentum scattering time due to pumping-induced heating
of the 2DES, which leads to an increase of the spin dephasing time in the motional-narrowing regime of the DP mechanism.
\subsection{Optical gating effects on the spin dephasing}
Finally, we study the influence of above-barrier illumination on the SDTs at low temperature.
\begin{figure}
  \includegraphics[width= 0.5 \textwidth]{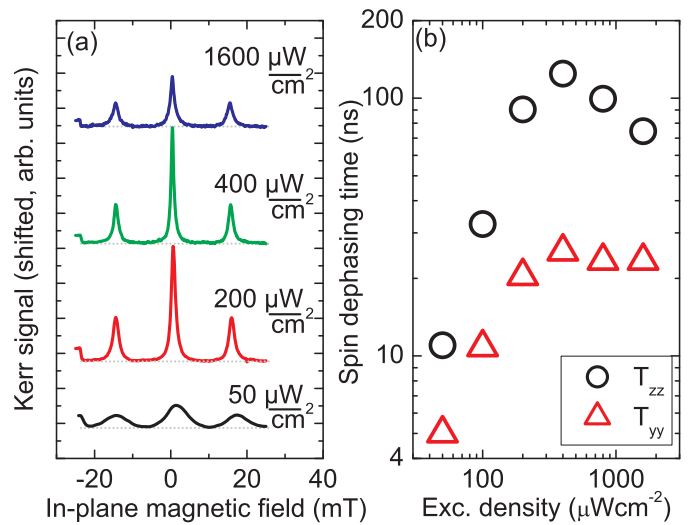}
   \caption{(a) RSA traces measured at  4~K for different excitation
     densities of above-barrier illumination. (b) Out-of plane
     ($T_{zz}$, open circles) and in-plane ($T_{yy}$, open triangles)
     spin dephasing times for different excitation densities of
     above-barrier illumination.}
   \label{Fig4}
\end{figure}
Figure \ref{Fig1}(d) shows the effect of weak, above-barrier
illumination on the 2DES PL: with increasing excitation density, the
width of the 2DES PL peak is reduced significantly, corresponding to a
partial depletion of the 2DES. This effect, which may even lead to the
inversion of the carrier type from $p$ to $n$ in a
$p$-modulation-doped QW~\cite{syperek07}, is often referred to
as optical gating and stems from a redistribution of charge carriers
from the 2DES to the remote dopant sites~\cite{Chaves1986}. This
reduction of the 2DES carrier density is directly visible in the RSA
traces in Figure \ref{Fig4}(a): for increasing illumination intensity,
both, the zero-field and finite-field peaks initially become more
pronounced and narrow, while they broaden again for higher
intensity. Correspondingly, the extracted SDTs
[Fig.~\ref{Fig4}(b)] drastically increase with illumination
intensity, reaching values above 150~ns for $T_{zz}$ and 25~ns for
$T_{yy}$, before decreasing again slightly for higher illumination
intensity. We may attribute this large increase to several effects:
the above-barrier illumination apparently symmetrizes the distribution of ionized
donors, thus reducing the growth-axis electric field and the
associated Rashba field. Additionally, the reduced carrier density
reduces the Fermi wave vector and thus, the  SOI for electrons at the
Fermi surface, slowing down spin dephasing due to the DP mechanism, as
well as reducing the single-electron momentum relaxation time due to
decreased Coulomb screening and increased electron-electron scattering
rate~\cite{Glazov04_eeI}. It is noteworthy, that even at the lowest
  possible electron densities we do not observe any decrease of the RSA
  peak amplitude with increasing the magnetic field, which rules out
  spin dephasing of localized electrons.  Therefore, we may infer that
  the observed decrease of   the SDTs for high above-barrier illumination intensity stems from
  the BAP mechanism which becomes relevant due to the increased hole
  density in the QW.
\section*{Conclusion}
In conclusion, we have investigated the spin dephasing in a
high-mobility (110)-grown two-dimensional electron system by resonant
spin amplification measurements. We observe a strong anisotropy of the
SDTs for in- and out-of-plane spin orientation, as well as a strong
temperature dependence of the out-of-plane SDT, which we attribute to
dopant-ionization-related changes in the growth-axis electric
field.  For weak above-barrier-illumination, SDTs above 150~ns
are reached at low temperatures for delocalized carriers,
exceeding previously reported values
for (110)-grown 2DES by an order of magnitude.

The authors would like to thank D.R. Yakovlev and S.A. Tarasenko for
fruitful discussion. Financial support  by the DFG via SFB689 and SPP
1285, RFBR, Dynasty Foundation, MCINN of Spain (grant
FIS2009-12773-C02-01), and Basque Country Government (grant IT-472-10) is gratefully acknowledged.

\bibliography{RSA110}
\end{document}